\documentclass[9pt,shortpaper,twoside,web]{IEEEtran}
\usepackage{cite}
\usepackage{amsmath,amssymb,amsfonts}
\usepackage{graphicx}

\newtheorem{secthm}{Theorem}[section]

\newtheorem{secprob}[secthm]{Problem}
\newtheorem{secdefn}[secthm]{Definition}
\newtheorem{secrem}[secthm]{Remark}

\newcommand{\bR} { {\mathbb R}}

\newcommand{\cC} { {\cal C}}
\newcommand{\cI} { {\cal I}}
\newcommand{\cJ} { {\cal J}}

\def\red{\hfill $\lhd$}

\allowdisplaybreaks[4]

\begin{document}
\title{Virtual Contraction Approach to Decentralized Adaptive Stabilization of Nonlinear Time-Delayed Networks}
\author{Yu Kawano and Zhiyong Sun
\thanks{The work of Yu Kawano was partially supported by JST FOREST Program Grant Number JPMJFR222E and JSPS KAKENHI Grant Number JP24K00910. Z. Sun was supported in part by ``The Fundamental Research Funds for the Central Universities, Peking University''. }
\thanks{Yu Kawano is with Graduate School of Advanced Science and Engineering, Hiroshima University, Higashi-Hiroshima 739-8527, Japan (e-mail: ykawano@hiroshima-u.ac.jp). }
\thanks{Zhiyong Sun is with State Key Laboratory for Turbulence and Complex Systems, Department of Mechanics and Engineering Science, College of Engineering, Peking University, Beijing 100871, China (e-mail: zhiyong.sun@pku.edu.cn).}}

\maketitle

\begin{abstract}
In this paper, we exploit a diagonally dominant structure for the decentralized stabilization of unknown nonlinear time-delayed networks. {To this end, we first introduce a novel generalization of virtual contraction analysis to diagonally dominant time-delayed control systems.} We then show that nonlinear time-delayed networks can be stabilized using diagonal high-gains, provided that the input matrices satisfy certain generalized (column/row) diagonally dominant conditions. To enable stabilization of unknown networks, we further propose a distributed adaptive tuning rule for each individual gain function, guaranteeing that all closed-loop trajectories converge to the origin while the gains converge to finite values. The effectiveness of the proposed decentralized adaptive control is illustrated through a case study on epidemic spreading control in SIS networks with transmission delays.
\end{abstract}



\section{Introduction}
In stabilizing an unknown plant, one approach is to incrementally increase a feedback gain until the closed-loop system becomes stable. This process, known as the \emph{high-gain adaptive stabilization} problem, involves the challenge of identifying a suitable gain. While a substantial body of research exists on this topic, these studies primarily focus on single plants. Recently, the paper~\cite{SRL:21} has studied decentralized high-gain adaptive stabilization for linear networks and demonstrated that a diagonally dominant structure is crucial when each local plant tunes its local gain based on its state information only. Following this line of research, we investigate decentralized high-gain adaptive stabilization for nonlinear time-delayed networks.


\subsubsection*{Literature Review}
For adaptive stabilization, there are plenty of researches in the literature including studies on adaptive control of nonlinear systems with a scalar high-gain function~\cite{KS:87}; adaptive stabilization of linearizable systems~\cite{SI:89}; adaptive stabilization of nonlinear systems with a scalar adaptive gain function~\cite{PBP:91}; adaptive tracking of nonlinear systems~\cite{AAI:97}; global adaptive stabilization with cascaded nonlinear systems~\cite{JSK:97}; adaptive nonlinear regulation \cite{PMI:06}; and adaptive designated-time stabilization of triangular nonlinear systems \cite{SLW:24}. Also, adaptive (high-gain) stabilization of nonlinear systems is closely related to passivity properties, as investigated by~\cite{DAlessandro:96}, and explored in works on feedback passive systems~\cite{SHF:95} and passivity-based adaptive stabilization~\cite{JHF:96}. We emphasize that these all mentioned papers focus on adaptive stabilization with scalar gain functions (i.e., a centralized adaptive scheme). {Thus, it is difficult to apply these adaptive tuning rules directly to decentralized control of networked systems.}

As aforementioned, the paper \cite{SRL:21} has studied decentralized adaptive stabilization of linear networks, identifying that a key structural property is diagonal dominance. This aligns with the well-known fact that diagonal dominance properties enable {decentralized} stability analysis of positive linear or monotone nonlinear systems \cite{HCH:10, Rantzer:15, RRD:13, FBJ:18, KBC:20, KC:22, JDB:23, KF:24, KH:25}. In the linear case, even with time delays, such results have been generalized to general linear systems (including positive systems) by taking component-wise absolute values of system matrices \cite{LA:80, ZZM:18}. Recent work on delay adaptive control has focused mostly on linear systems, as studied in the monograph \cite{zhu2020delay}. A diagonal dominance–based approach was explored in our preliminary work~\cite{KS:23} for decentralized adaptive stabilization of nonlinear delay-free networks using virtual contraction techniques \cite{LS:98,WS:05,JF:10,KC:22}. {However, incorporating time delays introduces additional challenges, as the Lyapunov theory for time-delayed systems~\cite{HL:13} differs from that for delay-free systems. Therefore, decentralized stabilization of nonlinear time-delayed networks remains an open problem.}


\subsubsection*{Contribution}
In this paper, we study decentralized high-gain adaptive stabilization of nonlinear time-delayed networks from the viewpoint of diagonal dominance. To this end, we first generalize virtual contraction analysis to time-delayed systems and establish sufficient conditions for global uniform asymptotic stability of a nonlinear time-delayed system, which identifies a diagonal dominance property by taking the component-wise absolute values of some Jacobian matrix. These conditions can be viewed as natural extensions of those previously derived conditions for \textit{linear} time-delayed systems in the classical work \cite{LA:80}. Utilizing one of them, we show that a \textit{nonlinear} time-delayed network with a diagonally dominant input matrix can be stabilized by selecting a gain of each node sufficiently large. In situations where system parameters are unknown, determining an appropriate gain amplitude for stabilization is very challenging, if not impossible. Therefore, we propose an automatic tuning rule for each gain, ensuring that all closed-loop trajectories converge to the origin without knowing system parameters while the gains converge to finite values. This constitutes a major contribution on the decentralized adaptive stabilization of nonlinear time-delayed networks. We also address the related dual problem arising in the context of decentralized output feedback control.

The contribution of this paper is summarized as follows:
\begin{itemize}
\item Generalizing virtual contraction analysis to time-delayed systems, we provide novel stability conditions for nonlinear time-delayed systems in terms of diagonal dominance;

\item For a nonlinear time-delayed network with a diagonally dominant input matrix, we propose a distributed automatic tuning rule for each local gain ensuring that all closed-loop trajectories converge to the origin, thus advancing the theory of decentralized adaptive stabilization for time-delayed uncertain networks. 
\end{itemize}


\subsubsection*{Organization}
The remainder of this paper is organized as follows. In Section~\ref{PF:sec}, we mention the studied problem of decentralized adaptive stabilization. In Section~\ref{sec:ASA}, we study stability of nonlinear time-delayed systems in terms of diagonal dominance. In Section~\ref{sec:DAS}, as the main result, we provide an adaptive tuning rule for decentralized high-gain stabilization of a nonlinear time-delayed network with a diagonally dominant input matrix.

\subsubsection*{Notation}
Let $\bR$ and $\bR_{>0}$ denote the sets of real numbers and positive real numbers, respectively. For a vector, its Euclidean norm is denoted by $| \cdot |$. Let $\cC := C( [-T_d, 0], \bR^n)$ denote the Banach space of continuous functions mapping the interval $[-T_d, 0]$ into $\bR^n$ with the topology of uniform convergence. For $\phi \in \cC$, define its norm by $\| \phi \|_\cC := \sup_{\theta \in [-T_d, 0]} |\phi (\theta)|$. For any $T>0$ and $x \in C( [-T_d, T], \bR^n)$, let $x_t \in \cC$, $t \in[0, T]$ denote a segment of the function $x$ defined by $x_t(\theta ) = x(t + \theta )$, $\theta \in [-T_d, 0]$. The $n \times n$ identity matrix is denoted by $I_n$. 


\section{Problem Formulation}\label{PF:sec}
In this paper, we are interested in generalizing results \cite{SRL:21} on decentralized adaptive stabilization for linear delay-free networks to nonlinear time-delayed networks, described by
\begin{align*}
&\dot x(t) = f(t, x(t), {x(t-T_1(t)),\dots, x(t-T_r(t))}) \\
&\quad \qquad + B(t, x(t), {x(t-T_1(t)),\dots, x(t-T_r(t))}) u(t), \; t \ge \sigma\\
&x(t+\sigma) = \phi (t), \;  t \in [-T_d, 0 ], \; {T_d := \max_{k=1,\dots,r}\{T_k(\sigma ) \}},
\end{align*}
where $\sigma > 0$, {$T_\ell: \bR \to \bR_{>0}$, $\ell=1,\dots,r$ such that $\max_{\ell=1,\dots,r}\{\dot T_\ell(t) \} \le d < 1$ for all $t \in \bR$}, and $\|\phi\|_\cC$ is bounded; $f : \bR \times \bR^n \times \bR^{rn} \to \bR^n$ is continuous and continuously differentiable in the second {to last} arguments and satisfies $f(t, 0, {0, \dots,} 0) = 0$ for all $t \in \bR$; $B: \bR \times \bR^n \times \bR^{rn} \times  \bR^n \to \bR^{n \times n}$ is continuous and locally Lipschitz in the second {to last} arguments, which represents the coupling term and decentralized control in network systems. 

By abuse of notation, each $x_i \in \bR$ denotes the state of node~$i$. As a control problem, we consider the scenario where each node designs its control input $u_i:\bR \to \bR$ only based on its \textit{local} information $x_i$ by $u_i = k_i(t) x_i$, $i=1,\dots,n$ for decentralized stabilization, whereas $f$ and $B$ are unknown. Utilizing diagonal gain matrix $K(t) := {\rm diag} \{ k_1(t), \dots, k_n(t) \}$, the closed-loop system can be represented as
\begin{align}\label{eq:stab}
\dot x(t) 
&= f(t, x(t), {x(t-T_1(t)),\dots, x(t-T_r(t))}) \nonumber\\
&\quad + B(t, x(t), {x(t-T_1(t)),\dots, x(t-T_r(t))}) K(t) x(t),
\end{align}
where $K :\bR \to \bR^{n \times n}$ is continuous.
Our main objective is as follows.

\begin{secprob}\label{prob:main}
For the time-delayed network~\eqref{eq:stab} with unknown $f$ and $B$, design a diagonal gain matrix $K$ such that all closed-loop trajectories converge to the origin.
\red
\end{secprob}

{It is worth mentioning that it is difficult to apply existing adaptive tuning rules, e.g., \cite{KS:87,SI:89,PBP:91,AAI:97,JSK:97,PMI:06,SLW:24,DAlessandro:96,SHF:95,JHF:96} directly to decentralized control of networked systems.}
In the linear delay-free case, it has been shown by~\cite[Theorem 4]{SRL:21} that Problem~\ref{prob:main} has a solution if the coupling matrix $B$ possesses a kind of diagonally dominant properties. We investigate its extension to nonlinear time-delayed networks
{
because diagonal dominance provides a tractable, scalable and decentralized way to analyze large-scale networks without requiring global information. In particular, for positive linear systems~\cite{HCH:10,Rantzer:15} and monotone nonlinear systems~\cite{KF:24,KH:25}, diagonal dominance enables stability analysis through decomposition into \textit{component-wise} problems. Also, from a practical viewpoint, diagonal dominance is reasonable in various settings where self-regulation or damping outweighs coupling effects. Examples include: (i) cooperative control of multi-agent systems, where strong local feedback ensures that interactions remain bounded; (ii) power networks with droop-controlled inverters, where local damping dominates coupling through transmission lines; and (iii) biological regulatory networks, where degradation rates of species often exceed the influence of cross-coupling terms.}


\section{Virtual Contraction Theory for\\ Diagonally Dominant Time-Delayed Nonlinear Systems}\label{sec:ASA}
To generalize results of the linear delay-free case~\cite[Theorem 4]{SRL:21}, we in this section study stability of nonlinear time-delayed systems in terms of diagonal dominance. Diagonal dominance properties have been utilized for stability analysis in a variety of systems, particularly for positive and monotone systems, e.g., \cite{HCH:10,Rantzer:15,FBJ:18,KBC:20,Kawano:20,KC:22,JDB:23,KF:24}, some of which has {its} root in system contraction theory \cite{FB-CTDS}.  A similar approach has been investigated for general linear time-delayed systems~\cite{LA:80,ZZM:18} by taking the component-wise absolute value of the system matrix. Inspired by this, we derive novel two stability conditions based on diagonally dominant types properties for nonlinear time-delayed systems. 


In this section, we proceed with stability analysis of a closed-loop system, described by
\begin{align}\label{eq:sys}
&\dot x = f(t, x(t), {x(t-T_1(t)),\dots, x(t-T_r(t))}), \; t \ge \sigma\\
&x(t+\sigma) = \phi (t), \;  t \in [-T_d, 0 ], \; {T_d := \max_{k=1,\dots,r}\{T_k(\sigma ) \}},
\nonumber
\end{align}
where we recall that $f(t, 0, {0, \dots,} 0) = 0$ for all $t \in \bR$.

We consider the following standard stability properties \cite[Definition 1.1]{HL:13}.
\begin{secdefn}
The solution $x=0$ of the system~\eqref{eq:sys} is said to be 
\begin{itemize}
\item \emph{uniformly stable} if 
for any $\sigma \in \bR$ and $\varepsilon >0$, there exists $\delta = \delta (\varepsilon) > 0$ such that 
\begin{align*}
\| \phi \|_\cC \le \delta
\;
\implies
\;
\| x_t \|_\cC \le \varepsilon,
\quad 
\forall t \ge \sigma;
\end{align*}

\item \emph{uniformly asymptotically stable} (UAS) if
it is uniformly stable, and there exists $\delta_0 > 0$ such that for every $\eta > 0$, there exists $t_0=t_0 (\delta_0,\eta)$ such that
\begin{align*}
\| \phi \|_\cC \le \delta_0
\;
\implies
\;
\| x_t \|_\cC \le \eta,
\quad
\forall t \ge \sigma + t_0
\end{align*}
for every $\sigma \in \bR$;

\item \emph{globally uniformly asymptotically stable} (GUAS) if
it is uniformly stable, and $\delta_0$ can be made arbitrary large. 
\red
\end{itemize}
\end{secdefn}

To analyze GUAS, we generalize a technique of \textit{virtual contraction} analysis \cite{LS:98,WS:05,JF:10,KC:22} to time-delayed systems. Let us introduce~$g: \bR \times \bR^n \times \bR^{rn} \times \bR^n \times \bR^{rn} \to \bR^n$ such that
\begin{subequations}\label{eq1:g}
\begin{align}
&g(t, x, {y_1,\dots,y_r}, x, {y_1,\dots,y_r}) = f(t, x, {y_1,\dots,y_r}) \\
&g(t, x, {y_1,\dots,y_r}, 0, {0,\dots, }0) = 0
\end{align}
\end{subequations}
for all $(t, x, {y_1,\dots,y_r}) \in \bR \times \bR^n \times \bR^{rn}$, where $g(t, x, {y_1,\dots,y_r}, \xi, {\eta_1,\dots,\eta_r})$ is continuous, locally Lipshitz in $x$ and ${y_1,\dots,y_r}$, and continuously differentiable in $\xi$ and $ {\eta_1,\dots,\eta_r}$. Then, $f$ can be represented by
\begin{align}\label{eq2:g}
&f(t, x, {y_1,\dots,y_r}) \nonumber\\
&=  g(t, x, {y_1,\dots,y_r}, x, {y_1,\dots,y_r}) \nonumber\\
&= \int_0^1 \frac{d g(t, x, {y_1,\dots,y_r}, sx, {sy_1,\dots,sy_r})}{ds} ds \nonumber\\
&= \int_0^1 \frac{\partial g(t, x, {y_1,\dots,y_r}, \xi, {\eta_1,\dots,\eta_r})}{\partial \xi} \biggl|_{(\xi, {\eta_\ell}) = (sx, {sy_\ell})} x ds \nonumber\\
&\quad + \int_0^1  {\sum_{\ell=1}^r}\frac{\partial g(t, x, {y_1,\dots,y_r}, \xi, {\eta_1,\dots,\eta_r})}{\partial {\eta_\ell}} \biggl|_{(\xi, {\eta_\ell}) = (sx, {sy_\ell})} {y_\ell} ds.
\end{align}
{In the delay-free case, stability analysis  of an equilibrium via contraction analysis of $g$ is refereed to as virtual contraction analysis, e.g., \cite{LS:98,WS:05,JF:10,KC:22}. For monotone delay-free systems, contraction analysis has been proceeded based on diagonal dominance, e.g., \cite{KBC:20,Kawano:20,KC:22,JDB:23,KF:24}. As the first main result of this paper, we further generalize this approach to virtual contraction analysis of not-necessarily monotone nonlinear systems with time delays by leveraging Lyapunov theory for time-delayed control systems~\cite{HL:13}. Our approach opens a new door to contraction analysis of time-delayed systems itself because specifying $g$ into $f$ yields (non-virtual) contraction conditions for time-delayed systems, although contraction analysis lies beyond the scope of this paper.
}

\begin{secthm}\label{thm:GUAS}
For a system~\eqref{eq:sys}, suppose that there exist $a \ge 0$ and $g: \bR \times \bR^n \times \bR^{rn} \times \bR^n \times \bR^{rn} \to \bR^n$ satisfying~\eqref{eq1:g} and 
\begin{align}\label{eq3:g}
 \left| \frac{\partial g(t, x, {y_1,\dots,y_r}, \xi, {\eta_1,\dots,\eta_r})}{\partial \eta_{\ell,j}} \right| \le a\\
\quad
\forall i,j = 1,\dots,n, \; \ell =1,\dots,r
\nonumber
\end{align}
for all $(t, x, {y_1,\dots,y_r}, \xi, {\eta_1,\dots,\eta_r}) \in \bR \times \bR^n \times \bR^{rn} \times \bR^n \times \bR^{rn}$, {where $\eta_{\ell,j}$ denote the $j$th component of $\eta_\ell$.} The solution $x=0$ to the system is GUAS if either of the following two conditions holds:

\begin{enumerate}
\renewcommand{\theenumi}{\Roman{enumi}}
\item there exist $c > 0$ and element-wise positive $v \in \bR^n$ such that
\begin{align}\label{eq:v}
&v_j \left(\frac{a{r}}{1-d} + \frac{\partial g_j (t, x, {y_1,\dots,y_r}, \xi, {\eta_1,\dots,\eta_r})}{\partial \xi_j} \right) \nonumber\\
&\quad + \sum_{i \neq j} v_i \left( \frac{a{r}}{1-d} +  \left|\frac{\partial g_i (t, x, {y_1,\dots,y_r}, \xi, {\eta_1,\dots,\eta_r})}{\partial \xi_j} \right|\right) \nonumber\\
&\le - c v_j
\end{align}
for all $j=1, \dots, n$ and $(t, x, {y_1,\dots,y_r}, \xi, {\eta_1,\dots,\eta_r}) \in \bR \times \bR^n \times \bR^{rn} \times \bR^n \times \bR^{rn}$;
\item there exist $c > 0$ and element-wise positive $w \in \bR^n$ such that
\begin{align}\label{eq:w}
&\left( a{r}+ \frac{\partial g_i (t, x, {y_1,\dots,y_r}, \xi, {\eta_1,\dots,\eta_r})}{\partial \xi_i} \right) w_i\nonumber\\
&\quad + \sum_{j \neq i} \left( a{r}+ \left|\frac{\partial g_i (t, x, {y_1,\dots,y_r}, \xi, {\eta_1,\dots,\eta_r})}{\partial \xi_j} \right|\right) w_j \nonumber\\
&\le - c w_i
\end{align}
for all $i=1, \dots, n$ and $(t, x, {y_1,\dots,y_r}, \xi, {\eta_1,\dots,\eta_r}) \in \bR \times \bR^n \times \bR^{rn} \times \bR^n \times \bR^{rn}$.
\end{enumerate}
\end{secthm}

\begin{IEEEproof}
The proof is in Appendix~\ref{app:GUAS}.
\end{IEEEproof}

In Theorem~\ref{thm:GUAS}, the conditions in items I) and II) can be understood as generalized diagonally-dominant type conditions of the Jacobian matrix $\partial g/\partial \xi$ {because we require that each diagonal term $|\partial g_i/\partial \xi_i|$, $i=1,\dots,n$ is sufficiently large compared with the (weighted) sum of the absolute values of corresponding off-diagonal elements in the same row and column.} The difference is that items I) and II) are with respect to the column and row sums, respectively, which correspond to the 1-norm and infinity-norm of relevant Lyapunov vector functions in virtual contraction analysis.  

{
\begin{secrem}
In the proof for item II) of Theorem~\ref{thm:GUAS}, we do not use $\max_{\ell=1,\dots,r}\{\dot T_\ell(t) \} \le d$. In fact, the statement can be shown for time-dependent delays $T_\ell (t)$, $\ell=1,\dots,r$ such that $T_\ell : \bR \to \bR$ is continuous and $\lim_{t \to \infty} t - T_\ell(t) = \infty$, by following the condition in \cite{FBJ:18}.  
\red
\end{secrem}
}


\section{Decentralized Adaptive Stabilization}\label{sec:DAS}
In this section, we apply item I) of Theorem~\ref{thm:GUAS} to solve Problem~\ref{prob:main}. We first show that if $B$ possesses a kind of diagonally dominant properties then there exist sufficiently large $k_i$, $i=1,\dots,n$, rendering the closed-loop system~\eqref{eq:stab} to be GUAS. Then, we provide an automatic tuning rule of each local gain $k_i$ for solving Problem~\ref{prob:main}, only utilizing its local information $x_i$. After that, we address the dual problem  of Problem~\ref{prob:main} based on item II) of Theorem~\ref{thm:GUAS}.


\subsection{Main Results}
We first characterize a class of matrices $B$ for which the system~\eqref{eq:stab} can be made GUAS by selecting $k_i$, $i=1,\dots,n$ sufficiently large, as the second main results of this paper.

\begin{secthm}\label{thm:DAS}
Suppose that the nonlinear time-delayed network system~\eqref{eq:stab} satisfies the following two conditions:
\begin{enumerate}
\renewcommand{\theenumi}{\Roman{enumi}}
\item $\partial f(t, x,  {y_1,\dots,y_r})/\partial x$, $\partial f(t, x,  {y_1,\dots,y_r})/\partial y_\ell$, {$\ell=1,\dots,\allowbreak r$} and $B(t,x, {y_1,\dots,y_r})$ are bounded on $(t, x,  {y_1,\dots,y_r}) \in \bR \times \bR^n \times \bR^{nr}$;

\item there exist $c > 0$ and element-wise positive $v \in \bR^n$ such that
\begin{align*}
&v_j B_{j,j} (t, x, {y_1,\dots,y_r}) 
+ \sum_{i \neq j} v_i | B_{i,j} (t, x, {y_1,\dots,y_r}) |\\
&\le - c v_j, \quad \forall j=1,\dots,n
\end{align*}
for all $(t, x, {y_1,\dots,y_r}) \in \bR \times \bR^n \times \bR^{nr}$.
\end{enumerate}
Then, there exist $k_i > 0$, $i=1,\dots,n$ such that the solution $x=0$ to the system is GUAS.
\end{secthm}

\begin{IEEEproof}
The proof is in Appendix~\ref{app:DAS}.
\end{IEEEproof}

Theorem~\ref{thm:DAS} shows that nonlinear delayed networks can be stabilized by selecting each local gain $k_i$, $i=1,\dots,n$ sufficiently large. However, their lower bounds are difficult or impossible to estimate if $f$ and $B$ are unknown. To resolve this issue, we provide an \textit{adaptive} tuning rule of each decentralized gain $k_i$, stated as the third main result of this paper. 

\begin{secthm}\label{thm2:DAS}
Suppose that the uncertain time-delayed  system~\eqref{eq:stab} satisfies items I) and II) of Theorem~\ref{thm:DAS}.
We implement the following update rule for each local gain $k_i(t)$, $i=1,\dots,n$:
\begin{align}\label{eq:gain}
\dot k_i(t) = \min\{ a_i, b_i |x_i(t-{T^k_i})| \}, \quad 
t \ge \sigma,
\end{align}
where {$T^k_i \in [0, T_d]$ and} $a_i, b_i > 0$ for all $i=1,\dots,n$.
Then, the following statements hold for all initial conditions $\sigma \in \bR$, $\phi \in \cC$, and $k_i(\sigma) \ge 0$ and for all $i=1,\dots,n$:

\begin{enumerate}
\renewcommand{\theenumi}{\Roman{enumi}}
\item $x_i(t)$ and $k_i(t)$ uniquely exist for all $\sigma \le t \in \bR$; 

\item $\lim_{t \to \infty} |x_i(t)| = 0$;

\item there exists $k_i^\infty > 0$ such that $\lim_{t \to \infty} k_i(t) = k_i^\infty$.
\end{enumerate}
\end{secthm}

\begin{IEEEproof}
The proof is in Appendix~\ref{app2:DAS}.
\end{IEEEproof}

{
In the delay-free case, Theorem~\ref{thm2:DAS} is equivalent to \cite[Theorem~5]{SRL:21} for linear systems and \cite[Theorem~8]{KS:23} for nonlinear systems. To achieve this generalization, we establish novel stability conditions for time-delayed systems in Theorem~\ref{thm:GUAS}. Moreover, we show the convergence of $k_i(t)$. In the linear case, this follows directly from exponential stability, and thus no additional proof is required, in contrast to the nonlinear case. In the delay-free nonlinear system setting, the convergence proof is not provided in \cite[Theorem~8]{KS:23}. Here, we provide such a proof and further show that the tuning rule~\eqref{eq:gain} can also be applied when using delayed state $x_i(t-T_i^k)$.  

In practice, each gain $k_i(t)$ may be subject to an allowable upper bound $\bar{k}_i$. However, it is generally difficult, or even impossible, to determine in advance whether such bounds are sufficiently large to guarantee stabilization for nonlinear control systems with unknown parameters. To account for this practical constraint, albeit without a theoretical guarantee, the adaptation law can be modified so that each $k_i(t)$ either stops updating or is reset to zero once it reaches its allowable upper bound $\bar{k}_i$.  
}


\subsection{Dual Problems}
In the linear case \cite{SRL:21}, decentralized adaptive stabilization has been also studied in the context of \textit{output} feedback. Its nonlinear generalization is
\begin{align}\label{eq:ob}
&\dot x(t) = f(t, x(t), {x(t-T_1),\dots, x(t-T_r)}) \nonumber\\
&\qquad + K(t) H(t, x(t), {x(t-T_1),\dots, x(t-T_r)}), \; t \ge \sigma \nonumber\\
&x(t+\sigma) = \phi (t), \;  t \in [-T_d, 0 ], \; {T_d := \max_{k=1,\dots,r}\{T_k(\sigma ) \}}
\end{align}
where $H: \bR \times \bR^n  \times \bR^{nr} \to \bR^n$ is continuous and continuously differentiable in the second and third arguments such that $H(t, 0, 0 {,\dots, 0}) = 0$ for all $t \in \bR$. Then, the origin is an equilibrium point.

Applying item II) of Theorem~\ref{thm:GUAS}, we obtain the counterpart of Theorem~\ref{thm:DAS} for high-gain decentralized stabilization.

\begin{secthm}\label{thm3:DAS}
Suppose that the nonlinear uncertain network system~\eqref{eq:ob} satisfies the following two conditions:
\begin{enumerate}
\renewcommand{\theenumi}{\Roman{enumi}}
\item $\partial f(t, x, {y_1,\dots,y_r})/\partial x$, $\partial f(t, x, {y_1,\dots,y_r})/\partial y_\ell$, $\partial H(t, x, {y_1,\dots,y_r})/\partial x$, and $\partial H(t, x, {y_1,\dots,y_r})/\partial y_\ell$, {$\ell =1,\dots,r$} are bounded on $(t, x, {y_1,\dots,y_r}) \in \bR \times \bR^n \times \bR^{nr}$;

\item there exist $c > 0$ and element-wise positive $w \in \bR^n$ such that
\begin{align*}
&\frac{\partial H_i (t, x, {y_1,\dots,y_r})}{\partial x_i} w_i  + \sum_{j \neq i}  \left| \frac{\partial H_i (t, x, {y_1,\dots,y_r})}{\partial x_j} \right| w_j\\
&\le - c w_i, \quad \forall i=1, \dots, n
\end{align*}
for all $(t, x, {y_1,\dots,y_r}) \in \bR \times \bR^n \times \bR^{nr}$.

\end{enumerate}
Then, there exist local gains $k_i > 0$, $i=1,\dots,n$ such that the solution $x=0$ to the closed-loop system is GUAS.
\end{secthm}

\begin{IEEEproof}
The proof is in Appendix~\ref{app3:DAS}.
\end{IEEEproof}

We also obtain the counterpart of Theorem~\ref{thm2:DAS} for high-gain decentralized adaptive stabilization as follows.

\begin{secthm}\label{thm4:DAS}
Suppose that the nonlinear uncertain network system~\eqref{eq:ob} satisfies items I) and II) of Theorem~\ref{thm3:DAS}.
We implement the update rule \eqref{eq:gain} for $k_i(t)$, $i=1,\dots,n$.
Then, items I) -- III) of Theorem~\ref{thm2:DAS} hold for all initial conditions $\sigma \in \bR$, $\phi \in \cC$, and $k_i(\sigma) \ge 0$ and for all $i=1,\dots,n$.
\end{secthm}

\begin{IEEEproof}
The proof is similar to that of Theorem~\ref{thm2:DAS}, and thus is omitted.
\end{IEEEproof}


\section{Example on  decentralized control of eliminating epidemic spreading in networks}
Recent years have witnessed increasing attention on the decentralized control of eliminating epidemic spreading in {delay-free} networks~\cite{walsh2025decentralised}. In this paper, we apply the developed decentralized adaptive approach to stabilizing a network of the Susceptible-Infected-Susceptible (SIS) model~\cite{MMZ:17} with delays, given by
\begin{align}\label{eq:SIS}
&\dot x_i (t) = (1 - x_i(t)) \sum_{j=1}^n c_{i,j} x_j(t - T) - k_i x_i(t), \; t \ge 0\nonumber\\
&x_i (t) = \phi_i (t), \;  t \in [-T, 0], 
\end{align}
where $x_i(t) \in [0, 1]$ denotes the fraction of infected individuals in population $i$ at time $t \ge 0$, and $T \ge 0$ represents the epidemic transmission delays. Also, $c_{i,j} \ge 0$, $i,j=1,\dots,n$ denotes the rate at which the infected individuals in population~$j$ can transmit the disease to the susceptibles in population~$i$. 

{In the recent work \cite{walsh2025decentralised}, decentralized adaptive-gain control is developed for delay-free epidemic networks. However, the results~\cite{walsh2025decentralised} for the delay-free epidemic network cannot directly handle transmission delays in the model or controller. In dynamical systems, time-delays are sometimes approximated by first-order lag systems (in this example, exposed dynamics). Its accuracy deteriorates when the delay is large relative to the system dynamics. To capture the effect of delays more precisely, we directly analyze the time-delayed system.}

In this section, we apply Theorem~\ref{thm2:DAS} for adaptive stabilization to tune the recovery parameter $k_i$, $i=1,\dots,n$ in each population~$i$ such that the epidemic disappears in network SIS systems with transmission delays. Note that the trajectory $x(t)$ stays in $[0, 1]^n$ for any $t \ge 0$ and initial condition $\phi(t) \in [0, 1]^n$, $t \in [-T, 0]$. {To show this, consider
\begin{align*}
\dot x_i (t) = (1 - x_i(t)) \sum_{j=1}^n c_{i,j} u_j(t) - k_i x_i(t), \; t \ge 0.
\end{align*}
When $u_j(t) = x_j(t-T)$, this becomes \eqref{eq:SIS}. We first show that $x_i(t) \ge 0$ for all $t \ge 0$ if $x_i(0) \ge 0$ and $u_j(t) \ge 0$, $j=1,\dots,1$ for all $t \ge 0$. Suppose that there exists $\tau \ge 0$ such that $x_i (\tau) = 0$. Then, we have 
\begin{align*}
\dot x_i (\tau) = \sum_{j=1}^n c_{i,j} u_j(\tau) \ge 0.
\end{align*}
This implies that $x_i(t)$ cannot be less than zero. Next, we show that $x_i(t) \le 1$ for all $t \ge 0$ if $x_i(0) \le 1$ and $u_j(t) \in [0, 1]$, $j=1,\dots,n$ for all $t \ge 0$. Suppose that there exists $\tau \ge 0$ such that $x_i (\tau) = 1$. Then, we have $\dot x_i (\tau) = - k_i x_i(\tau)$.
This implies that $x_i(t)$ cannot be greater than one. Therefore, if $\phi_i(t) \in [0,1]$, $i=1,\dots,n$ for all $t \in [-T, 0]$, then it holds that $x_i(t) \in [0,1]$, $i=1,\dots,n$ for all $t \ge 0$.}

On the bounded set $[0, 1]^n$, item I) of Theorem~\ref{thm:DAS} automatically holds. Moreover, item II) holds because $B$ in the representation~\eqref{eq:stab} is $B : = -I_n$ in this case. Therefore, by implementing decentralized adaptive tuning rule~\eqref{eq:gain}, the state trajectories converge to the origin. Moreover, each gain converges to a finite value.

\begin{figure}[t]
  \centering
  \includegraphics[width = 0.85\linewidth]{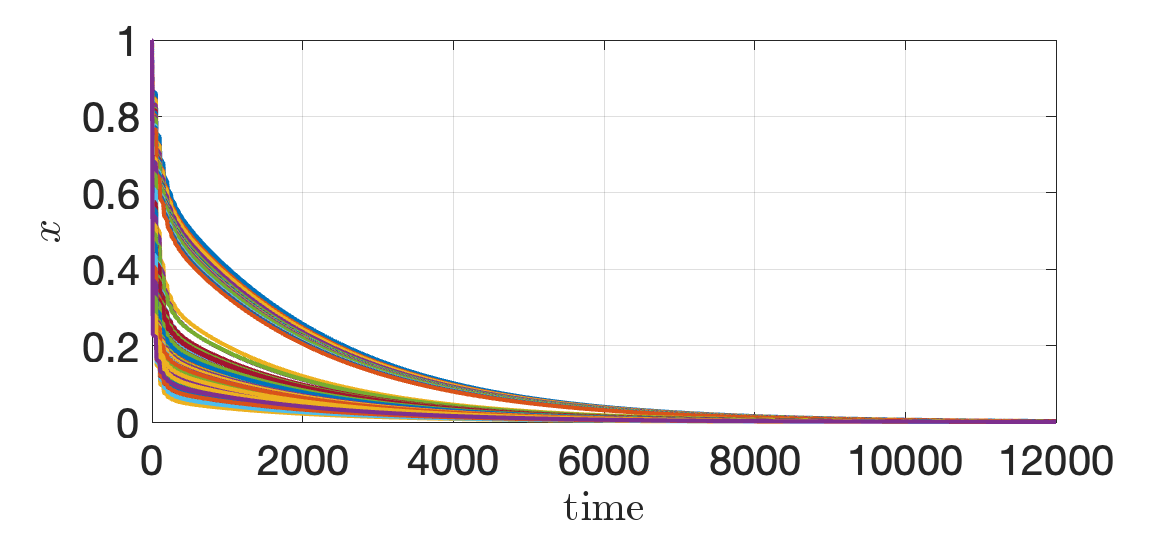}
  \caption{Time response of the SIS network with decentralized adaptive control law, where all trajectories converge to zero. }
  \label{fig:simx}
\vspace{4mm}
  \includegraphics[width = 0.85\linewidth]{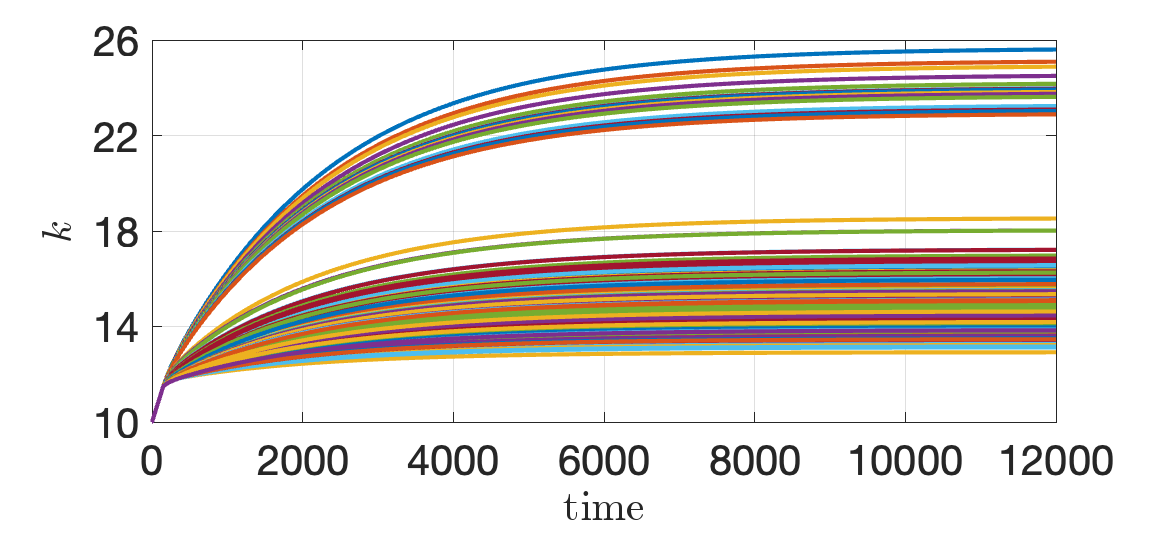}
  \caption{Time response of the decentralized adaptive gains}
  \label{fig:simk}
\end{figure}

\begin{figure}[!t]
  \centering
  \includegraphics[width = 0.85\linewidth]{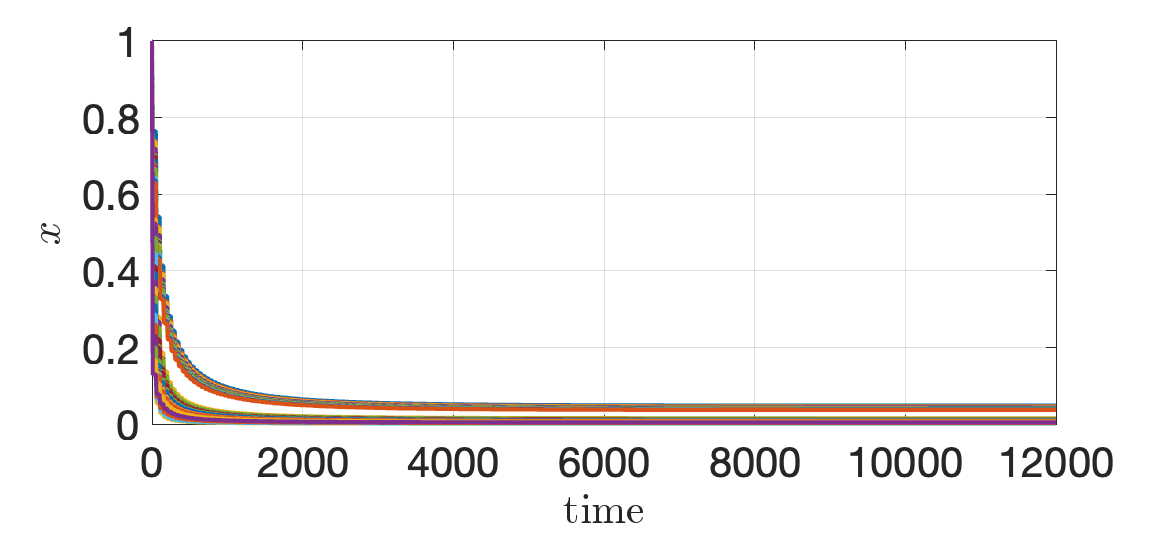}
  \caption{Time response of the SIS network when $k_i=20$, where certain trajectories do not converge to zero.}
  \label{fig:simk20}
\end{figure}

To conduct numerical simulation, we select the time delay as $T = 50$.  For the network topology, we consider a {randomly generated} scale-free network with $n = 200$ nodes and $5$ edges per each node. 
{
{Regarding the controller parameters, there is an inevitable delay in obtaining infection information for determining the amount of vaccination. To capture this effect, we set $T_i^k = 100$ for all $i = 1, \dots, n$.}
Also, we select $a_i = 0.1$, $b_i = 0.01$, and $k_i(0) = 10$ for all $i = 1, \dots, n$. 

As confirmed by Figs.~\ref{fig:simx} and~\ref{fig:simk}, the proposed method achieves decentralized adaptive stabilization, and each gain $k_i(t)$ converges to a finite value with the average being approximately $15.2$. To further illustrate the utility of the proposed adaptive tuning rule, Fig.~\ref{fig:simk20} presents the system trajectories with constant local gains $k_i(t) \equiv 20 > 15.2$, $i = 1, \dots, n$. In this case, some trajectories fail to converge to zero, indicating that epidemic spreading in networks cannot be eliminated by conventional decentralized control with fixed gains. These results highlight the effectiveness and necessity of the proposed adaptive strategy.
}


\section{Conclusion}
In this paper, we have studied decentralized adaptive stabilization for unknown nonlinear time-delayed networks with diagonally dominant input matrices. First, we have generalized diagonally dominant-type stability conditions from positive linear systems to nonlinear time-delayed systems which are not required to be positive or monotone. In this way, we have contributed to the development of virtual contraction analysis to time-delayed systems.  Based on the column diagonally dominant condition, we have clarified the structure of diagonally dominant input matrices, enabling the stabilization of nonlinear time-delayed networks using decentralized high-gain control. Additionally, we have provided an automatic tuning rule for each gain in a decentralized manner, without the need for any system parameter knowledge. Furthermore, we have investigated the dual of decentralized adaptive stabilization by utilizing the row diagonally dominant condition. The proposed decentralized adaptive control method has been applied for eliminating epidemic spreading in SIS networks with transmission delays.


\appendices
\section{Proof of Theorem~\ref{thm:GUAS}}\label{app:GUAS}
(item I)
Consider the following Lyapunov function candidate
\begin{align}\label{pf1:ES}
V_s (x(t)) := \sum_{i=1}^n v_i \left( |x_i(t)| + {\frac{a}{1-d} \sum_{\ell=1}^r \int_{t-T_\ell(t)}^t} \sum_{j=1}^n |x_j(\theta)| d\theta \right).
\end{align}
From~\eqref{eq2:g} and $\max_{i=1,\dots,r}\{\dot T_i(\cdot) \} \le d < 1$,
The upper right Dini derivative of $V_s(x)$ along the trajectory of the time-delayed nonlinear system~\eqref{eq:stab}, denoted by $D^+ V_s$, satisfies
\begin{align*}
&D^+ V_s (x(t)) \\
&= \sum_{i=1}^n v_i \frac{x_i(t)}{|x_i(t)|} f_i(t, x(t), {x(t-T_1),\dots, x(t-T_r)}) \\
&\quad + {\frac{a}{1-d} \sum_{\ell=1}^r} \sum_{i=1}^n \sum_{j=1}^n v_i (|x_j (t)| - {(1- \dot T_\ell)}|x_j(t-{T_\ell})|) \\
&\le \int_0^1 \sum_{i=1}^n \sum_{j=1}^n v_i \frac{x_i(t)}{|x_i(t)|}  \biggr(\frac{\partial g_i}{\partial \xi_j} x_j(t) \\
&\qquad  \qquad  + {\sum_{\ell=1}^r} \frac{\partial g_i}{\partial {\eta_{\ell,j}}} x_j(t-{T_\ell})
\biggl)\biggl|_{(\xi, {\eta_\ell}) = (sx, {sx(t-{T_\ell})})} ds\\
&\quad +  a {\sum_{\ell=1}^r} \sum_{i=1}^n \sum_{j=1}^n v_i \left(\frac{|x_j (t)|}{1-d} - |x_j(t-{T_\ell})|\right),
\end{align*}
where the arguments $(t, x(t), {x(t-T_1(t)),\dots, x(t-T_r(t))}, \xi, {\eta_1,}$
${\dots,\eta_r})$ of $g_i$ and $t$ of $T_\ell(t)$ are dropped for the notational simplicity. It follows from $|x_j(t)| = \int_0^1|x_j(t)| ds$ that
\begin{align}\label{pf1-1:ES}
D^+ V_s (x(t)) 
&\le \int_0^1 \sum_{j=1}^n A_j ds + \int_0^1 B  ds \nonumber\\
&\qquad - a {\sum_{\ell=1}^r} \sum_{i=1}^n \sum_{j=1}^n v_i |x_j(t-{T_\ell})|
\end{align}
with
\begin{align*}
A_j &:= \sum_{i=1}^n v_i \biggl( \frac{ar}{1-d} |x_j(t)| +  \frac{\partial g_i}{\partial \xi_j} \frac{x_i(t) x_j(t)}{|x_i(t)|} \biggr)
\end{align*}
and
\begin{align*}
B := {\sum_{\ell=1}^r}  \sum_{i=1}^n \sum_{j=1}^n v_i \frac{\partial g_i}{\partial {\eta_{\ell,j}}} \frac{x_i(t) x_j (t-{T_\ell})}{|x_i(t)|},
\end{align*}
where the substitution $(\xi, {\eta_\ell}) = (sx, {sx(t-{T_\ell})})$ in $g_i$ is dropped  for the notational simplicity.

From~\eqref{eq:v}, we have
\begin{align*}
A_j 
&= v_j\left( \frac{ar}{1-d} + \frac{\partial g_j}{\partial \xi_j} \right) |x_j|  
+\sum_{i \neq j} v_i \biggl(\frac{ar}{1-d} |x_j| +  \frac{\partial g_i}{\partial \xi_j} \frac{x_i x_j}{|x_i|} \biggr) \\
&\le \left( v_j\left( \frac{ar}{1-d} + \frac{\partial g_j}{\partial \xi_j} \right)   
+ \sum_{i \neq j}  v_i \biggl(\frac{ar}{1-d}  +  \left|\frac{\partial g_i}{\partial \xi_j}\right|  \biggr) \right) |x_j|\\
&\le - c v_j |x_j|.
\end{align*}
Also, from~\eqref{eq3:g}, we obtain
\begin{align*}
B 
&\le {\sum_{\ell=1}^r}  \sum_{i=1}^n \sum_{j=1}^n v_i \left|\frac{\partial g_i}{\partial {\eta_{\ell,j}}}\right| \frac{|x_i(t)| |x_j (t-{T_\ell})|}{|x_i(t)|}\\
&\le a {\sum_{\ell=1}^r}  \sum_{i=1}^n \sum_{j=1}^n v_i |x_j (t-{T_\ell})|.
\end{align*}
Combining them with~\eqref{pf1-1:ES} leads to
\begin{align*}
D^+ V_s (x(t)) \le - \sum_{i=1}^n  c v_i |x_i(t)|.
\end{align*}
By Lyapunov-Krasovskii Theorem, e.g.~\cite[Theorem 1]{Fridman:14}, the solution $x=0$ is GUAS.

(item II)
Define 
\begin{align}\label{pf2:ES}
V_m (x) := \max_{i=1,\dots,n} \left\{ \frac{|x_i|}{w_i}\right\}.
\end{align}
For this function, we define the following set of the indexes:
\begin{align*}
I(x) := \left\{ i=1, \dots, n : \frac{|x_i|}{w_i} = V_m(x) \right\}.
\end{align*}
From its definition, we have
\begin{align}\label{pf3:ES}
\frac{|x_j|}{w_j}
\le
\frac{|x_i|}{w_i}
\quad
\iff
\quad
|x_j| \le \frac{w_j}{w_i} |x_i|
\end{align}
for all $j=1,\dots, n$ and $i \in I(x)$.

It follows from~\eqref{eq2:g} that
\begin{align*}
&D^+ V_m (x(t)) \\
&= \max_{i \in I(x(t))} \frac{x_i(t)}{w_i |x_i(t)|} f_i(t, x(t), {x(t-T_1),\dots, x(t-T_r)}) \\
&= \max_{i \in I(x(t))} \frac{x_i(t)}{w_i |x_i(t)|}
\int_0^1 \sum_{j=1}^n \biggr(\frac{\partial g_i}{\partial \xi_j} x_j(t) \\
&\qquad  \qquad  + {\sum_{\ell=1}^r} \frac{\partial g_i}{\partial {\eta_{\ell,j}}} x_j(t-{T_\ell})
\biggl)\biggl|_{(\xi, {\eta_\ell}) = (sx, {sx(t-{T_\ell})})}  ds,
\end{align*}
where the arguments $(t, x(t), {x(t-T_1(t)),\dots, x(t-T_r(t))}, \xi, {\eta_1,}$
${\dots,\eta_r})$ of $g_i$ and $t$ of $T_\ell(t)$ are dropped for the notational simplicity.
The first term of the most right-hand side can be upper bounded as
\begin{align*}
\frac{x_i}{w_i|x_i|} \sum_{j=1}^n \frac{\partial g_i}{\partial \xi_j} x_j
&= \frac{x_i}{w_i|x_i|} 
\left( \frac{\partial g_i}{\partial \xi_i} x_i
+ \sum_{j \neq i} \frac{\partial g_i}{\partial \xi_j} x_j \right)\\
&= \frac{1}{w_i} 
\left( \frac{\partial g_i}{\partial \xi_i} |x_i|
+ \sum_{j \neq i} \frac{\partial g_i}{\partial \xi_j} \frac{x_j x_i}{|x_i|}\right)\\
&\le \frac{1}{w_i} 
\left( \frac{\partial g_i}{\partial \xi_i} |x_i|
+ \sum_{j \neq i} \left| \frac{\partial g_i}{\partial \xi_j} \right| |x_j| \right).
\end{align*}
Also, it follows from~\eqref{eq:w} and~\eqref{pf3:ES} that
\begin{align*}
&\max_{i \in I(x)} \frac{1}{w_i} 
\left( \frac{\partial g_i}{\partial \xi_i} |x_i|
+ \sum_{j \neq i} \left| \frac{\partial g_i}{\partial \xi_j} \right| |x_j| \right)\\
&\le \max_{i \in I(x)}  \frac{|x_i|}{w_i^2} 
\left( \frac{\partial g_i}{\partial \xi_i} w_i
+ \sum_{j \neq i} \left| \frac{\partial g_i}{\partial \xi_j} \right| w_j  \right)\\
&\le \max_{i \in I(x)}  \frac{|x_i|}{w_i^2} 
\left(- (c + a{r}) w_i - \sum_{j \neq i} a{r} w_j\right)\\
&= - c  V_m (x) - \max_{i \in I(x)} \sum_{j = 1}^n \frac{a{r} w_j}{w_i} V_m (x).
\end{align*}
Next, from~\eqref{eq3:g}, the second term satisfies
\begin{align*}
&\max_{i \in I(x(t))} \frac{x_i(t)}{w_i |x_i(t)|} \sum_{j=1}^n {\sum_{\ell=1}^r} \frac{\partial g_i}{\partial {\eta_{\ell,j}}} x_j(t-{T_\ell})\\
&\le \max_{i \in I(x(t))} \frac{|x_i(t)|}{w_i |x_i(t)|} \sum_{j=1}^n {\sum_{\ell=1}^r} \left|\frac{\partial g_i}{\partial {\eta_{\ell,j}}} \right| |x_j(t-{T_\ell})|\\
&\le \max_{i \in I(x(t))} \frac{a}{w_i} \sum_{j=1}^n {\sum_{\ell=1}^r} |x_j(t-{T_\ell})|\\
&\le \max_{i \in I(x(t))} \sum_{j=1}^n \frac{a w_j}{w_i} {\sum_{\ell=1}^r} V_m (t-{T_\ell}).
\end{align*}
In summary, we have
\begin{align*}
D^+ V_m (x(t)) 
&\le - c V_m (x(t)) \\
&+ \max_{i \in I(x(t))}  \sum_{j=1}^n \frac{a w_j}{w_i} {\sum_{\ell=1}^r} (V_m (t-{T_\ell(t)}) - V_m (x(t)) ).
\end{align*}
To apply Lyapunov-Razumikhin Theorem, e.g.~\cite[Theorem 2]{Fridman:14}, suppose that {$V_m(x(t-T_\ell(t))) \le (1+\varepsilon) V_m(x(t))$, $\ell=1,\dots,r$} for some $\varepsilon > 0$. Then, it follows that
\begin{align*}
D^+ V_m (x(t)) 
&\le  \left(  - c + \varepsilon {r} \max_{i \in I(x)} \sum_{j=1}^n \frac{a w_j}{w_i} \right) V_m(x(t)).
\end{align*}
For $\varepsilon > 0$ satisfying $\varepsilon  < c/ {r}(\max_{i =1,\dots,n} \sum_{j=1}^n a w_j/w_i)$, we have $\bar c := c - \varepsilon {r} \max_{i \in I(x)} \sum_{j=1}^n a w_j/w_i > 0$. Therefore, we obtain
\begin{align*}
D^+ V_m (x(t)) \le  - \bar c V_m(x(t)).
\end{align*}
By Lyapunov-Razumikhin Theorem, the solution $x=0$ is GUAS.
\IEEEQED


\section{Proof of Theorem~\ref{thm:DAS}}\label{app:DAS}
For the system, we choose $g$ as
\begin{align}\label{pf1:DAS}
&g (t, x, {y_1,\dots,y_r}, \xi, {\eta_1,\dots,\eta_r}) \nonumber\\
&= f(t, \xi, {\eta_1,\dots,\eta_r})  + B(t, x, {y_1,\dots,y_r}) K \xi,
\end{align}
which satisfies \eqref{eq1:g}.
For sufficiently large $k_i > 0$, $i=1,\dots,n$, we show that $g$ satisfies~\eqref{eq:v}. Then, the statement of this theorem follows from item I) of Theorem~\ref{thm:GUAS}.

First, from item I) of this theorem, there exists $a \ge 0$ such that the condition of \eqref{eq3:g} and the following inequality hold
\begin{align*}
v_j \frac{\partial f_j (t,\xi,{\eta_1,\dots,\eta_r})}{\partial \xi_j} 
+ \sum_{i \neq j} v_i \left|\frac{\partial f_i(t,\xi, {\eta_1,\dots,\eta_r})}{\partial \xi_j}\right|
\le \frac{ar}{1-d} v_j \; \\
\forall j=1,\dots,n, \; 
\forall (t, \xi, {\eta_1,\dots,\eta_r}) \in \bR \times \bR^n  \times \bR^{nr}.
\nonumber
\end{align*}
Next, item II) of this theorem implies 
\begin{align*}
& v_j B_{j,j}  (t, x, {y_1,\dots,y_r}) k_j
+ \sum_{i \neq j} v_i | B_{i,j}  (t, x, {y_1,\dots,y_r}) | k_j \\
&\le - c v_j k_j
\le - c \min_{j=1,\dots,n}\{k_j\}  v_j \\
&\quad \forall j=1,\dots,n, \; 
(t, x, {y_1,\dots,y_r}) \in \bR \times \bR^n \times \bR^n {\times \cdots \times \bR^n}.
\end{align*}
Thus, it follows from the selection~\eqref{pf1:DAS} of $g$ that
\begin{align}\label{pf2:DAS}
&v_j \left(\frac{ar}{1-d} + \frac{\partial g_j (t, x, {y_1,\dots,y_r}, \xi, {\eta_1,\dots,\eta_r})}{\partial \xi_j} \right) \nonumber\\
&\quad + \sum_{i \neq j} v_i \left( \frac{ar}{1-d}+  \left|\frac{\partial g_i (t, x, {y_1,\dots,y_r}, \xi, {\eta_1,\dots,\eta_r})}{\partial \xi_j} \right|\right) \nonumber\\
&\le \left( \frac{ar}{1-d} \left( 1 + \sum_{i=1}^n \frac{v_i}{v_j} \right) - c \min_{j=1,\dots,n}\{k_j\}\right)  v_j.
\end{align}
There exist sufficiently large $k_j > 0$, $j=1,\dots,n$ such that
\begin{align*}
\frac{ar}{1-d} \left( 1 + \sum_{i=1}^n \frac{v_i}{v_j} \right) - c \min_{j=1,\dots,n}\{k_j\} < 0,
\quad
\forall j=1,\dots,n.
\end{align*}
That completes the proof.
\IEEEQED


\section{Proof of Theorem~\ref{thm2:DAS}}\label{app2:DAS}

(Proof of item I)) 
It follows that
\begin{align*}
&B(t,x,{y_1,\dots,y_r}) K x - B(t,x',{y'_1,\dots,y'_r}) K' x'\\
&= B(t,x,{y_1,\dots,y_r}) K x - B(t,x,{y_1,\dots,y_r}) K x' \\
&\quad + B(t,x,{y_1,\dots,y_r}) K x' - B(t,x,{y_1,\dots,y_r}) K' x'\\
&\quad + B(t,x,{y_1,\dots,y_r}) K' x' - B(t,x',{y'_1,\dots,y'_r}) K' x',
\end{align*}
and consequently
\begin{align*}
&|f(t,x,{y_1,\dots,y_r}) + B(t,x,{y_1,\dots,y_r}) K x \\
&\quad - f(t,x',{y'_1,\dots,y'_r}) - B(t,x',{y'_1,\dots,y'_r}) K' x'|\\
&\le |f(t,x,{y_1,\dots,y_r}) - f(t,x',{y'_1,\dots,y'_r})| \\
&\qquad + |B(t,x,{y_1,\dots,y_r}) K| |x - x'| \\
&\quad + |B(t,x,{y_1,\dots,y_r})| |K - K'| |x'| \\
&\quad + |B(t,x,{y_1,\dots,y_r}) - B(t,x',{y'_1,\dots,y'_r})| | K' x'|.
\end{align*}
Recall that $f$ is continuous and continuously differentiable in $(x, {y_1,\dots,y_r})$, and $B$ is continuous and locally Lipschitz in $(x, {y_1,\dots,y_r})$.
Thus, $f + B K x$ is continuous and locally Lipschitz in $(x, {y_1,\dots,y_r}, K)$.
Moreover, $|x_i|$ is locally Lipschitz in $x_i$.
Therefore, $x_i (t)$ and $k_i(t)$, $i=1,\dots,n$ exist in some time interval. Also, they are unique as long as they exist; see e.g. \cite[Theorem 2.3]{HL:13}.

From~\eqref{pf2:DAS}, $V_s(x)$ in~\eqref{pf1:ES} satisfies
\begin{align*}
&D^+ V_s(x(t)) \le \left( \hat a - c \min_{j=1,\dots,n}\{k_j (t)\}\right) \sum_{i=1}^n v_i |x_i(t)|\\
&\quad \hat a := \frac{ar}{1-d} \left( 1 + \sum_{i=1}^n \frac{v_i}{\min_{j=1,\dots,n} \{ v_j \}} \right).
\end{align*}
By~\eqref{eq:gain}, each $k_i(t)>0$, $i=1,\dots,n$ is an increasing function of time, and thus, from~\eqref{pf1:ES},
\begin{align*}
&D^+ V_s(x(t)) \le \hat b(\sigma) \sum_{i=1}^n v_i |x_i(t)| \le \hat b(\sigma) V_s (x(t))\\
&\qquad \hat b(\sigma) := \hat a - c \min_{j=1,\dots,n}\{k_j(\sigma)\}.
\end{align*}
The comparison principle \cite[Lemma 3.4]{Khalil:96} and again~\eqref{pf1:ES} lead to
\begin{align}\label{pf0:DAS2}
v_i |x_i(t)| 
\le V_s(x(t))
\le V_s(x(\sigma)) e^{ \hat b(\sigma) (t - \sigma)} 
\end{align}
for all $i=1,\dots,n$. Thus, $x_i(t)$, $i = 1,\dots,n$ stay in $\bR$ for all $\sigma \le t \in \bR$, $\phi \in \cC$, and $k_i(\sigma) \ge 0$, $i = 1,\dots,n$.

Also, it follows from~\eqref{eq:gain} and~\eqref{pf0:DAS2} that
\begin{align}
&k_i(t) \nonumber\\
&= k_i(\sigma) + \int_\sigma^t \min\{a_i, b_i |x_i (\tau-{T^k_i})|\} d\tau \label{pf00:DAS2}\\
&\le k_i(\sigma) + \int_\sigma^{\sigma + {T^k_i}} \min\{a_i, b_i |x_i (\tau-{T^k_i})|\} d\tau\nonumber\\
&\quad 
+ \int_{\sigma + {T^k_i}}^t \min\{a_i, b_i |x_i (\tau-{T^k_i})|\} d\tau\nonumber\\
&= k_i(\sigma) + \int_\sigma^{\sigma + {T^k_i}} \min\{a_i, b_i |x_i (\tau-{T^k_i})|\} d\tau\nonumber\\
&\quad 
+  \min\left\{a_i (t - \sigma-{T^k_i} ), \frac{b_i V_s(x(\sigma))}{v_i \hat b(\sigma)}  \left(e^{ \hat b(\sigma) (t - \sigma -{T^k_i})} - 1 \right)\right\}\nonumber
\end{align}
for all $t \ge \sigma + {T^k_i}$. {In the interval $[\sigma, \sigma + {T^k_i}]$, it is clear that $k_i(t)$ is bounded.} 
Thus, $k_i(t)$, $i = 1,\dots,n$ stay in $\bR$ for all $\sigma \le t \in \bR$, $\phi \in \cC$, and $k_i(\sigma) \ge 0$, $i = 1,\dots,n$. Namely, neither $x_i(t)$ or $k_i(t)$ has a finite escape time.

(Proofs of items II) and III)) 
We divide $i=1,\dots,n$ into two subsets $\cI, \cJ \subset \{1,\dots,n\}$.
First, $\cI$ is the set of $i$ for which there exists $\sigma \le t_i \in \bR$ such that
\begin{align}\label{pf1:DAS2}
\hat a - \hat c k_i(t_i) < 0.
\end{align}
Second, $\cJ$ is the set of $j$ for which such $t_j$ does not exists.
In other words, all $j \in \cJ$ satisfy
\begin{align}\label{pf2:DAS2}
k_j(t) = k_j(\sigma) + \int_\sigma^t \min\{a_j, b_j |x_j (\tau {-T^k_j})|\} d\tau \le  \frac{\hat a}{\hat c} \; \\
\quad 
\forall t \ge \sigma.
\nonumber
\end{align}
From their definitions, it holds that $\cI \cap \cJ = \emptyset$ and $\cI \cup \cJ = \{1,\dots,n\}$.

We first consider $j \in \cJ$. Since the integral in~\eqref{pf2:DAS2} is an upper bounded increasing sequence of $t \ge \sigma$, this converges to a finite value, i.e., item III) holds for all $j \in \cJ$.
In addition, there exists a sufficiently large $(\sigma \le) t_j \in \bR$ such that $b_j |x_j(t)| \le a_j$ for all $t \ge t_j$. That is, $|x_j(t)|$, $t \ge t_j$ is uniformly continuous. From Barbalat's Lemma, e.g., \cite[Lemma 8.2]{Khalil:96}, we have $\lim_{t \to \infty} |x_j(t)| = 0$, i.e., item II) holds for all $j \in \cJ$.

We next consider $i \in \cI$.
Define 
\begin{align}\label{pf3:DAS2}
\bar V_s (x(t)) := \sum_{i \in \cI} v_i \left( |x_i(t)| + {\frac{a}{1-d} \sum_{\ell=1}^r \int_{t-T_\ell(t)}^t} \sum_{j=1}^n |x_j(\theta)| d\theta \right).
\end{align}
Similar calculations as the proofs of Theorems~\ref{thm:GUAS} and \ref{thm:DAS} lead  to
\begin{align*}
&D^+ \bar V_s(x(t))\\
&\le \left( \hat a - \hat c \min_{i \in \cI}\{k_i (t)\} \right) \sum_{i\in \cI} v_i|x_i(t)| \\
&\quad +  \sum_{i \in \cI} \sum_{j \in \cJ} v_i \Biggl(\int_0^1 \left|\frac{\partial f_i(t, \xi, {\eta_1,\dots,\eta_r})}{\partial \xi_j}\right|_{(\xi,{\eta_\ell})=(sx(t), s {x(t-T_\ell)})} ds \\
&\quad  + \frac{ar}{1-d} + k_j(t) |B_{i,j}(t, x(t), {x(t-T_1),\dots, x(t-T_r)})| \Biggr) |x_j(t)|.
\end{align*}
Recall that $\partial f(t, \xi, {\eta_1,\dots,\eta_r})/\partial \xi$ and $B(t, x, {y_1,\dots,y_r})$ are bounded, and all $k_j(t)$, $j \in \cJ$ are monotonically increasing and satisfy item III).
Thus, there exists $0< \hat d \in \bR$ such that
\begin{align*}
& \sum_{i \in \cI} \sum_{j \in \cJ} v_i \Biggl(\int_0^1 \left|\frac{\partial f_i(t, \xi, {\eta_1,\dots,\eta_r})}{\partial \xi_j}\right|_{(\xi,{\eta_\ell})=(sx(t), s {x(t-T_\ell)})} ds \\
&\quad  + \frac{ar}{1-d} + k_j(t) |B_{i,j}(t, x(t), {x(t-T_1),\dots, x(t-T_r)})| \Biggr) |x_j(t)|\\
&\le \hat d \sum_{j \in \cJ} |x_j(t)|.
\end{align*}
In summary, we have
\begin{align*}
D^+ \bar V_s(x(t)) 
&\le \left( \hat a - \hat c \min_{i \in \cI}\{k_i (t)\} \right) \sum_{i\in \cI} v_i|x_i(t)| 
+ \hat d \sum_{j \in \cJ} |x_j(t)|.
\end{align*}
From the definition of $\cI$ (recall \eqref{pf1:DAS2}), there exist $\lambda_\cI > 0$ and $t_\cI \ge \sigma$ such that
\begin{align*}
D^+ \bar V_s(x(t)) 
\le - \lambda_\cI \sum_{i\in \cI} v_i|x_i(t)|
 + \hat d \sum_{j \in \cJ} |x_j(t)|, \quad
\forall t \ge t_\cI.
\end{align*}
Taking the time integration in the interval $[t_\cI, t]$ yields
\begin{align*}
&\bar V_s(x(t)) + \lambda_\cI \int_{t_\cI}^t \sum_{i\in \cI} v_i|x_i(\tau)| d\tau \\
&\le \bar V_s(x(t_\cI)) + \hat d \sum_{j \in \cI} \int_{t_\cI}^t | x_j(\tau)| d \tau,
\quad
\forall t \ge t_\cI.
\end{align*}
From~\eqref{pf2:DAS2} with $\lim_{t \to \infty} x_j(t) = 0$ for all $j \in \cJ$ and~\eqref{pf3:DAS2}, there exists $\bar d (t_\cI, x(t_\cI )) > 0$ such that
\begin{align*}
|x_i(t)|
+ \lambda_\cI \int_{t_\cI}^t |x_i(\tau)| d\tau
\le \bar d (t_\cI, x(t_\cI )) ,
\quad
 \forall t \ge t_\cI
\end{align*}
for all $i \in \cI$. Thus, $|x_i(t)|$ is uniformly continuous, and by Barbalat's Lemma, item II) holds for all $i \in \cI$. This further implies that there exists $t_i > t_0$ such that $b_i |x_i(t)| \le a_i$ for all $t \ge t_i$.
Finally, since $ \int_{t_\cI}^t |x_i(\tau)| d\tau$ is a bounded function of $(t_\cI \le) t \in \bR$, $k_i$ in~\eqref{pf00:DAS2} satisfies item III) for all $i \in \cI$.
\IEEEQED


\section{Proof of Theorem~\ref{thm3:DAS}}\label{app3:DAS}
For the system~\eqref{eq:ob}, we choose $g$ as
\begin{align}
&g (t , x, {y_1,\dots,y_r}, \xi, {\eta_1,\dots,\eta_r})\nonumber\\
&= f(t, \xi, {\eta_1,\dots,\eta_r}) + K H(t, \xi, {\eta_1,\dots,\eta_r}).
\end{align}
which satisfies \eqref{eq1:g}.

First, from item I), there exists $a \ge 0$ such that \eqref{eq3:g} and 
\begin{align*}
&\frac{\partial f_i(t, \xi, {\eta_1,\dots,\eta_r})}{\partial \xi_i}  w_i
+ \sum_{j \neq i}  \left|\frac{\partial f_i(t, \xi, {\eta_1,\dots,\eta_r})}{\partial \xi_j}\right| w_j\\
&\le a {r} w_i, \quad \forall j=1,\dots,n
\end{align*}
for all $(t, \xi, {\eta_1,\dots,\eta_r}) \in \bR \times \bR^n \times \bR^{nr}$.
Next, it follows from item II) that 
\begin{align*}
&k_i \frac{\partial H_i (t, \xi, {\eta_1,\dots,\eta_r})}{\partial \xi_i} w_i  + k_i \sum_{j \neq i}  \left| \frac{\partial H_i (t, \xi, {\eta_1,\dots,\eta_r})}{\partial \xi_j} \right| w_j \\
&\le - c k_i w_i \le - c \min_{i=1,\dots,n}\{k_i\} w_i
\end{align*}
for all $(t, \xi, {\eta_1,\dots,\eta_r}) \in \bR \times \bR^n \times \bR^{nr}$.
Therefore, from the definition~\eqref{pf1:DAS} of $g$, we have, for all $i=1,\dots,n$,
\begin{align*}
&\left( a{r}+ \frac{\partial g_i (t, x, {y_1,\dots,y_r}, \xi, {\eta_1,\dots,\eta_r})}{\partial \xi_i} \right) w_i\\
&\quad + \sum_{j \neq i} \left( a{r}+ \left|\frac{\partial g_i (t, x, {y_1,\dots,y_r}, \xi, {\eta_1,\dots,\eta_r})}{\partial \xi_j} \right|\right) w_j\\
&\le \left( a{r} \left( 1 + \sum_{j=1}^n \frac{w_j}{w_i} \right) - c \min_{i=1,\dots,n}\{k_i\} \right)  w_i.
\end{align*}
There exist sufficiently large $k_i > 0$, $i=1,\dots,n$ such that
\begin{align*}
a {r}\left( 1 + \sum_{j=1}^n \frac{w_j}{w_i} \right) - c \min_{i=1,\dots,n}\{k_i\} < 0,
\quad
\forall i=1,\dots,n.
\end{align*}
That completes the proof.
\IEEEQED


\bibliographystyle{IEEEtran}
\bibliography{ref}
\end{document}